\documentclass[aps,twocolumn,showpacs]{revtex4-2}

\usepackage{amssymb}
\usepackage{graphicx}
\usepackage{dcolumn}
\usepackage{bm}
\usepackage{amsmath}
\usepackage{soul,color}
\usepackage{textcomp}
\usepackage{times}

\usepackage[colorlinks,linkcolor=red,citecolor=blue]{hyperref}

\begin{document}

\title{Dynamical generation of geometric squeezing in interacting Bose-Einstein condensates}
\author{Li Chen$^{1}$}
\email{lchen@sxu.edu.cn}
\author{Fei Zhu$^{1}$}
\author{Zheng Tang$^{1}$}
\author{Liang Zeng$^{1}$}
\author{Jae Joon Lee$^{2}$}
\author{Han Pu$^{2}$}

\affiliation{
$^1${Institute of Theoretical Physics, State Key Laboratory of Quantum Optics and Quantum Optics Devices, Shanxi University, Taiyuan 030006, China}\\
$^2${Department of Physics and Astronomy, and Smalley-Curl Institute, Rice University, Houston, TX 77005, USA}
}

\begin{abstract}
When the rotating frequency of a non-interacting Bose-Einstein condensate (BEC) confined in a weak anisotropic harmonic potential is suddenly quenched to its trapping frequency, the condensate evolves from its ground state to a single-mode squeezed state with exponentially growing quantum fluctuation anisotropy. Such a squeezed state is called the geometrically squeezed state. However, for interacting BECs with two-body collisions, a similar quench only results in quantum fluctuations oscillating periodically without squeezing. In this work, we identify superfluid stability as the key factor behind this non-squeezing phenomenon, with the periodic oscillations arising from collective excitations of a stable collective excitation mode. By strategically breaking the stability criteria, we propose a dynamical approach for generating squeezing that can exponentially suppress quantum fluctuations in a relatively short time, surpassing the efficiency of existing experimental preparation schemes.
\end{abstract}

\maketitle

\section{Introduction}
The quantum Hall effect represents a key phenomenon in quantum physics \cite{Stern2008,Yoshioka2002,Tong2016,Klitzing2020}, offering both precise standards of electrical resistance and insights into topological states of matter with potential applications in quantum information science. Rotating Bose-Einstein Condensates (BECs) serve as an ideal platform to simulate and explore quantum Hall physics \cite{Fetter2009, Fetter2007, Madison2000, Shaeer2001, Schweikhard2004, Bretin2004, Mukherjee2022, Ho2001, Recati2001, Sonin2005, Sinha2005}, as the Coriolis force experienced by rotating neutral atoms mimics the Lorentz force on charged particles, effectively creating a synthetic gauge field. In the rotating reference frame, the atomic dynamics is separated into two independent modes: the guiding-center mode with energy $\hbar(\omega - \Omega)$ and the cyclotron mode with energy $\hbar(\omega + \Omega)$, where $\omega$ and $\Omega$ represent the harmonic trapping frequency and rotational angular velocity, respectively. At the \textit{critical velocity} ($\Omega = \omega$), the guiding-center energy vanishes, creating degenerate states analogous to Landau levels in electronic systems. This quantum-Hall regime promises rich physical phenomena, including vortex lattice melting \cite{Viefers2008, Cooper2001, Baym2004, Sinova2002}, and the emergence of the Yrast spectrum \cite{Mottelson1999} and bosonic quantum Hall states \cite{Regnault2003, Regnault2004, Furukawa2012, Regnault2013}.

Recent studies \cite{Fletcher2021, Sharma2022, Crépel2024, Chen2025}, inspired by the seminal experimental work \cite{Fletcher2021}, have demonstrated the possibility of preparing a novel quantum state in rotating BECs, called the \textit{geometric squeezed state}, which is based on the following principle.
For a non-interacting BEC rotating at the critical speed, an slightly anisotropic confining potential can induce a single-mode squeezing Hamiltonian $\sim \zeta [b^2 + (b^\dagger)^2]/2$ where $b$ ($b^\dag$) is the annihilation (creation) operator for the guiding-center mode, which satisfies the SU(1,1) algebra similar to the parametric conversion processes in quantum optics  \cite{Wu1986, Wu1987, Scully1997, Walls2008}, with the squeezing parameter $\zeta$ proportional to the small trap anisotropy $\varepsilon$. 
Hence when the rotating frequency $\Omega$ is quenched from 0 to the critical value $\omega$, the non-interacting BEC evolves from the ground state of the harmonic oscillator to a squeezed state, where one quadrature fluctuation in the guiding-center phase space is exponentially suppressed at the expense of the conjugate quadrature fluctuation which is exponentially amplified.
In real space, the BEC's density distribution evolves from a nearly isotropic profile to a strongly anisotropic Gaussian wave packet with a minimum width of $\sigma_\text{LLL}$, with $\sigma_\text{LLL}$ being the characteristic length of the lowest Landau level.

However, the squeezing behavior is dramatically different for interacting BECs in the Thomas-Fermi regime, where interaction energy far exceeds kinetic energy. Our numerical results (detailed in Sec.~\ref{StPre}) show that, at critical rotational speed, the isotropic initial state cannot be effectively squeezed, instead exhibiting near-equilibrium oscillations. In fact, in the geometric-squeezing experiment \cite{Fletcher2021} using an interacting BEC, a specially designed ramping curve $\Omega(t)$ was employed [see Fig.~\ref{Fig1}(a)] to overcome these oscillations and achieve effective squeezing. However, unlike the non-interacting case, the squeezing factor does not decay exponentially over time and is heavily dependent on the ramp details of $\Omega(t)$. 
These observations seemingly suggest that the dynamical preparation of geometric squeezing for interacting and non-interacting systems stems from different mechanisms, which motivates us to seek the underlying mechanism of the squeezing dynamics.

In this study, we elucidate the manner in which interactions modify the BEC's stability serves as a key factor for the phenomenon mentioned above.
Our study provides two key findings: 1) For interacting BECs, there exists an abnormal branch of steady states whose quadrupole oscillations are responsible for the aforementioned periodic oscillations and non-squeezing behaviors; 2) By properly breaking the stability criteria, we find a parameter regime that enables rapid squeezing of interacting BECs, avoiding the near-equilibrium oscillations. This method can achieve an exponential scaling of quantum fluctuation squeezing in a short time, and also allow the BEC to enter the lowest Landau level in the long time. Therefore, the quench scheme for squeezing generation in both non-interacting and interacting BECs can be described within a unified theoretical framework, with the key point being the breaking of the BEC's dynamical stability.

The rest of the paper is organized as follows: 
Sec.~\ref{Model} introduces the model of the rotating BEC and the concept of geometric squeezing. 
Sec.~\ref{StPre} discusses the dynamics of squeezing in interacting BECs, comparing the cases of ramping $\Omega$ and quenching $\Omega$. The former has been used in the experiment, whereas the latter corresponds directly to the single-particle case, but fails to achieve geometric squeezing.
In Sec.~\ref{Hydrodynamics}, we explain the failure of the quench approach in the context of hydrodynamics, and examine the details of collective oscillations. 
Sec.~\ref{RPS} is devoted to our quench approach for interacting BECs. 
Sec.~\ref{Discussion} discusses the rigidification of the BEC during the squeezing process and the considerations for experiments. 
A concise conclusion can be found in Sec.~\ref{Conclusion}.

\section{Single-Particle Geometric Squeezing} \label{Model}
We consider a BEC rotating along the $z$ axis. A tight confinement on the $z$ direction is applied such that the axial wave function can be effectively integrated out. Hence, in the lab frame, the single-particle Hamiltonian in the $x$-$y$ plane can be expressed as (setting $\hbar = 1$) 
\begin{equation}
h_0(t) = \frac{\mathbf{p}^2}{2m} + V(\mathbf{r},t),
\end{equation}
where $\mathbf{r} = (x,y)$ and $\mathbf{p} = (p_x,p_y)$ are the two-dimensional coordinates the momenta, respectively, and $m$ denotes the atomic mass. The external potential
$V(\mathbf{r},t) = V_0(R^{-1}(t)\cdot\mathbf{r})$
is time-dependent with
\begin{equation}
R(t) = 
\begin{pmatrix}
\cos \Omega t & -\sin \Omega t \\
\sin \Omega t & \cos \Omega t
\end{pmatrix}
\end{equation}
being the SO(2) rotational matrix, which effectuates a clockwise rotation of the static trapping potential $V_0(\mathbf{r})$ about the origin, characterized by the angular frequency $\Omega$. For convenience, one usually employs the rotating reference frame in which the Hamiltonian ${h}_0$ is time-independent. Specifically, ${h}_0$ in the rotating frame is related to ${h}(t)$ in the lab frame through the transformation: ${h}_0 = {U} {h}(t)  {U}^\dagger + i \dot{U} U^\dagger$, where ${U}(t) = \exp(i\Omega t {L}_z)$ is a time-dependent unitary transformation. Therefore, in the rotating frame, we have
\begin{equation}
h_0 = \frac{\mathbf{p}^2}{2m} + V_0(\mathbf{r}) - \Omega L_z,
\end{equation}
with ${L}_z = \mathbf{r}\times \mathbf{p} =  -i(x\partial_y - y\partial_x)$ being the $z$-component angular momentum. In the following discussion, we will stick to the rotating frame.

For an isotropic harmonic potential $V_0(\mathbf{r}) = \frac{1}{2}m\omega^2 r^2$, the Hamiltonian can be rewritten as
\begin{equation}
{h}_0 = \frac{({\mathbf{p}} - \mathbf{A})^2}{2m}  + V_0^\text{eff}(\mathbf{r}),
\label{h0_2}
\end{equation}
where $V_0^\text{eff} =  \frac{1}{2}m(\omega^2-\Omega^2)r^2$ is the modified effective trapping potential and $\mathbf{A} = (-m \Omega y, m \Omega x,0)^T$ is an effective gauge vector potential. The first term shows the analogy between rotating atoms and charged particles in a magnetic field $\mathbf{B} = \nabla \times \mathbf{A} = 2 m \Omega \hat{z}$. At critical rotation ($\Omega = \omega$), the effective trapping potential vanishes, and the energy spectrum exhibits Landau-level structures.

The system can be described using two independent modes: the \textit{cyclotron mode}
\begin{equation}
a = \frac{\xi + i \eta}{\sqrt {2 l_B^2}}, \ \ \ 
a^\dagger = \frac{\xi - i \eta}{\sqrt {2 l_B^2}},
\label{opa}
\end{equation}
with quadratures 
\begin{align}
\xi = \frac{x}{2} - \frac{p_y}{2 m \omega}, \quad \eta = \frac{y}{2} + \frac{p_x}{2 m \omega}; \label{xieta}
\end{align}
and the \textit{guiding-center} mode
\begin{equation}
b = \frac{X - i Y}{\sqrt {2 l_B^2}}, \ \ \ 
b^\dagger = \frac{X + i Y}{\sqrt {2 l_B^2}},
\label{opb}
\end{equation}
with quadratures 
\begin{equation}
X = \frac{x}{2} + \frac{p_y}{2 m \omega}, \quad Y = \frac{y}{2} - \frac{p_x}{2 m \omega}.
\label{XY}
\end{equation}
 Each of these two modes satisfies canonical bosonic commutation relations
\begin{equation}
\begin{aligned}
[a,a^\dagger] &= [b,b^\dagger]=1,\\
[\xi,\eta] &= i l_B^2,\\
[X,Y] &= -i l_B^2,
\end{aligned}
\label{commutator1}
\end{equation}
where $l_B = 1/\sqrt{2m\omega}$ is the magnetic length. The two modes commute with each other, i.e., 
\begin{equation}
\begin{aligned}
[a,b] &= [a,b^\dagger]=0,\\
[\xi,X] &= [\xi,Y] = [\eta,X] = [\eta,Y] = 0.
\end{aligned}
\label{commutator2}
\end{equation}
In terms of these modes, the Hamiltonian takes the form
\begin{equation}
\begin{aligned}
h_0 &=M \omega\left[\omega_+\left(\xi^2+\eta^2\right)+\omega_-\left(X^2+Y^2\right)\right] \\
&= \omega_+\left(a^{\dagger} a+\frac{1}{2}\right)+\omega_-\left(b^{\dagger} b+\frac{1}{2}\right),
\end{aligned}
\label{Hamiltonian3}
\end{equation}
where $\omega_\pm = \omega\pm\Omega$. This formulation has the following implications: (1) The system comprises two independent modes with energies $\omega_\pm$; (2) At critical rotation ($\Omega=\omega$), $\omega_- = 0$ such that the energy spectrum becomes standard Landau levels, with eigenstates $|n_a = 0, n_b\rangle$ called the lowest Landau level (LLL); (3) The cyclotron and guiding-center modes each span a phase space with conjugate variables governed by uncertainty relations due to their non-zero commutators [Eq.~(\ref{commutator1})].

The geometric squeezing can be realized by introducing a weak anisotropy into the system \cite{Fletcher2021}, such that the trapping potential is modified as
\begin{equation}
V_0(\mathbf{r}) = \frac{m \omega^2}{2} (x^2+y^2) + \frac{\varepsilon m \omega^2}{2} (x^2-y^2),
\label{V0}
\end{equation}
where $\varepsilon$ is a small dimensionless parameter quantifying the anisotropy. In this case, the harmonic trapping frequencies along the $x$- and $y$-axis become $\omega_x = \sqrt{1+\varepsilon}\omega$ and $\omega_y = \sqrt{1-\varepsilon}\omega$, respectively. Under an additional unitary transformation $G = \exp(-i\kappa m \omega x y)$, with $\kappa = \varepsilon \omega/2\Omega$, the single-particle Hamiltonian transforms into
\begin{equation}
\begin{aligned}
\tilde{h}_0=&\frac{p^2}{2 m}+\frac{1}{2} m \omega^2\left(1+\kappa^2\right)\left(x^2+y^2\right) \\
&+\kappa \omega\left(x p_y+y p_x\right)-\Omega\left(x p_y-y p_x\right).
\end{aligned}
\label{Hamiltonian4}
\end{equation}
By redefining the distorted coordinate and momentum operators as
\begin{equation}
\begin{aligned}
\tilde x &= (1+\kappa^2)^{1/4} x, \ \ \ 
\tilde p_x = (1+\kappa^2)^{-1/4} p_x, \\
\tilde y &= (1+\kappa^2)^{1/4} y, \ \ \ 
\tilde p_y = (1+\kappa^2)^{-1/4} p_y,
\end{aligned}
\end{equation}
one can similarly recast $\tilde{h}_0$ in terms of distorted cyclotron and guiding-center operators as
\begin{equation}
\begin{aligned}
\tilde h_0 =&\tilde \omega_+\left(\tilde{a}^{\dagger} \tilde{a}+\frac{1}{2}\right)+\tilde\omega_-\left(\tilde{b}^{\dagger} \tilde{b}+\frac{1}{2}\right) \\
&-\frac{ \zeta}{2}\left(\tilde{a}^{\dagger} \tilde{a}^{\dagger}+\tilde{a} \tilde{a}-\tilde{b}^{\dagger} \tilde{b}^{\dagger}-\tilde{b} \tilde{b}\right),
\end{aligned}
\label{Hamiltonian5}
\end{equation}
where $\zeta = \kappa \omega$ and 
\begin{equation}
\begin{aligned}
\tilde\omega_\pm = \sqrt{1+\kappa^2}\omega \pm \Omega.
\end{aligned}
\end{equation} 
Here $\tilde a$ and $\tilde b$ are defined analogously to Eqs.~(\ref{opa}) and (\ref{opb}), but incorporate the coordinates and momenta with tildes. One can also define the cyclotron quadratures ($\tilde \xi$ and $\tilde \eta$) and the guiding-center quadratures ($\tilde X$ and $\tilde Y$) analogous to Eqs.~(\ref{xieta}) and (\ref{XY}). In this case, all the commutators displayed in Eqs.~(\ref{commutator1}) and (\ref{commutator2}) remain unchanged for the operators with tilde. 

The second line of $\tilde{h}_0$ in Eq.~(\ref{Hamiltonian5}), induced by the trap anisotropy, underlies the foundation for geometric squeezing. For small $\kappa$, $\tilde x \approx x$, $\tilde y\approx y$, and $\tilde \omega_\pm \approx \omega_\pm$. At critical rotation, $\tilde \omega_+ \approx 2\omega$ and $\tilde \omega_- \approx 0$, which leads to the evolution operator for the guiding-center $\tilde b$-mode taking the form of a squeezing operator
$U_b(t) = \exp \left\{\frac{\zeta t}{2} [\tilde{b'}^2 - (\tilde b'^\dagger)^2] \right\},$
with $\tilde b' = e^{i \varphi} \tilde b$, $\varphi = -\pi/4$ being the squeezing angle. $U_b$ allows the guiding center mode to evolve from the ground state to the single-mode geometrically squeezed state, i.e., 
\begin{equation}
|0,S(t)\rangle = U_b(t)\, |0,0\rangle.
\label{squeezed_state}
\end{equation}
During the squeezing process, quantum fluctuations in the $\tilde{X}$-$\tilde{Y}$ phase space deform exponentially, i.e.,
\begin{equation}
\begin{aligned}
\Delta_\text{min}(t) &= \sqrt{ \langle \tilde X^2_\text{min} \rangle - \langle \tilde X_\text{min}\rangle^2 }= \Delta_\text{SQL} e^{- \zeta t}, \\
\Delta_\text{max}(t) &= \sqrt{ \langle \tilde Y^2_\text{max} \rangle - \langle \tilde Y_\text{max}\rangle^2 }= \Delta_\text{SQL} e^{ \zeta t},
\label{Deltat}
\end{aligned}
\end{equation}
where $\Delta_\text{min}$ and $\Delta_\text{max}$ respectively denote the minimal and maximal quantum fluctuations in the guiding-center  $\tilde X$-$\tilde Y$ phase space, and
\begin{equation}
	\Delta_\text{SQL} = \frac{l_B}{\sqrt{2}},
\end{equation}
is the standard quantum limit (SQL) arising from the commutators Eq.~(\ref{commutator1}).

It is also worth mentioning that, although the cyclotron mode in Eq.~(\ref{Hamiltonian5}) also possesses terms $\tilde{a}^2$ and $(\tilde a^\dagger)^2$, the predominant free evolution $\tilde{a}^\dagger \tilde{a}$ with rate $\tilde{\omega}_+ \gg \kappa\omega$ causes a rapid rotation of the phase space, preventing the generation of squeezing for the cyclotron mode. However, this difficulty can be overcome by a Floquet technique as recently proposed \cite{Chen2025}.

\section{Dynamics for Interacting BECs} \label{StPre}

Thus far, our discussion has focused on the single-particle picture, which is applicable to noninteracting BECs. For a BEC subject to two-body interactions, the system is described by the second quantized Hamiltonian
\begin{equation}
H = \int d^2 r {\psi}^{\dagger}(\mathbf{r}) h_0 {\psi}(\mathbf{r}) + \frac{g}{2} {\psi}^{\dagger}(\mathbf{r}){\psi}^{\dagger}(\mathbf{r}){\psi}(\mathbf{r}) {\psi}(\mathbf{r}),
\end{equation}
where ${\psi}(\mathbf{r})$ denote the bosonic field operator, $g = \sqrt{8\pi \omega_z /m} a_s$ characterizes the effective interaction strength in 2D, with $a_s$ the 3D $s$-wave scattering length, and $\omega_z$ the axial trapping frequency. In the mean-field framework, one can solve for the ground state and dynamical evolution of the system by propagating the Gross-Pitaevskii (GP) equation 
\begin{equation}
i \dot \psi=\left[-\frac{\nabla^2}{2 m} + V_0 + i \Omega\left(x \frac{\partial}{\partial y}-y \frac{\partial}{\partial x}\right)+g|\psi|^2\right] \psi
\label{GPE}
\end{equation}
in imaginary time and real time, respectively, with $V_0$ given in Eq.~(\ref{V0}).
 
\begin{figure}[t]
	 \includegraphics[width=0.48\textwidth]{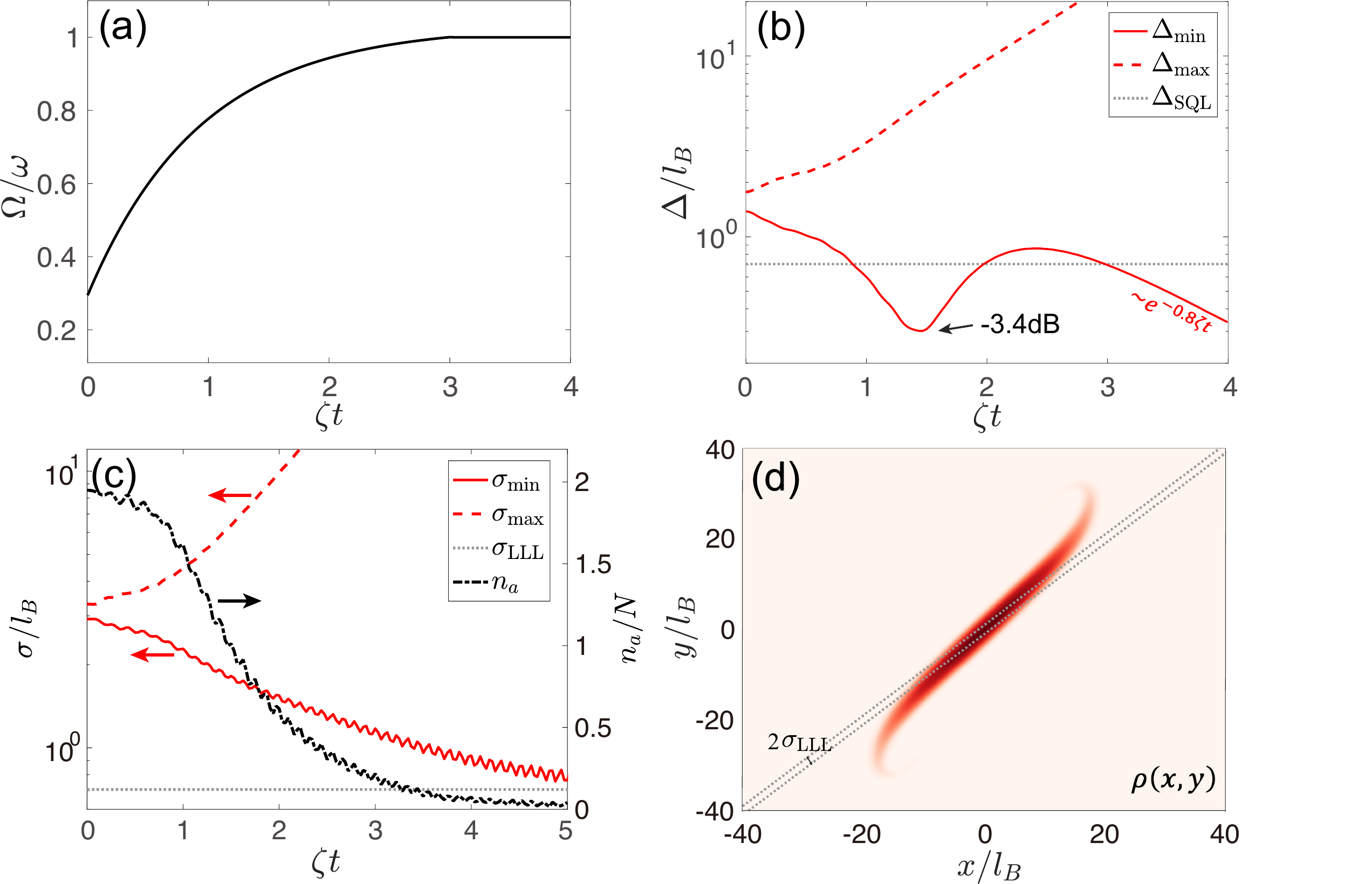}
	\caption{Suqeezing dynamics by ramping $\Omega$. (a) The ramping curve of $\Omega$. (b) Variations of the minimal $\Delta_\text{min}$ (solid line) and maximal quantum fluctuation $\Delta_\text{max}$ (dashed line) in the $\tilde X$-$\tilde Y$ phase space over time, where the standard quantum limit $\Delta_\text{SQL} = l_B/\sqrt{2}$ is shown by the dotted line. (c) Evolution of the minimal $\sigma_\text{min}$ (solid line) and maximal $\sigma_\text{max}$ (dashed line) wave-packet width in real space; the dotted line denotes $\sigma_\text{LLL} = l_B/\sqrt{2}$; the Landau-level occupation $n_a$ over time (dot-dashed line). Arrows indicate the corresponding axes for quantities. (d) The density distribution $\rho(x,y)$ at $\zeta t = 2.5$. The dotted line indicates $2\sigma_\text{LLL}$. In the calculation, $\varepsilon = 0.125$ is fixed, which leads to $\zeta = 0.0625 \omega$.}
	\label{Fig1}
\end{figure}

In the experiment \cite{Fletcher2021}, the geometric squeezed state was achieved using a fixed anisotropy parameter $\varepsilon = 0.125$ by initializing the BEC in a non-rotating ground state, then ramping $\Omega$ from zero to the critical value $\Omega=\omega$, and maintaining this value before measurement. We simulate this process by numerically solving the GP equation with parameters similar to the experiment: $N=5\times 10^4$ Na atoms in a harmonic trap with transverse frequency $\omega=88.6\times(2\pi)$ Hz and axial frequency $\omega_z=\sqrt{8} \omega$. The system has a Thomas-Fermi radius of approximately 11.2 $\mu$m, central trap density of $6\times 10^{13}$ cm$^{-3}$, and healing length of 443 nm. Fig.~\ref{Fig1}(a) shows the ramping curve for $\Omega(t)$; Figs.~\ref{Fig1}(b) and (c) display the evolution of quantum fluctuation $\Delta$ in the phase space of the guiding-center mode and the width of the wavepacket in real space $\sigma$, respectively.

In contrast to the single-particle case [Eqs.~(\ref{Deltat})] where $\Delta_\text{min}$ decays exponentially, the minimal quantum fluctuation shown in Fig.~\ref{Fig1}(b) behaves in a non-monotonic two-stage decay. 
$\Delta_\text{min}$ initially decreases to a minimum value at $\zeta t \approx 1.4$, reaching a squeezing parameter of about $10 \log_{10}(\Delta_\text{min}/\Delta_\text{SQL})\approx -3.4$dB, and then starts to rebound. Over longer timescales, $\Delta_\text{min}$ decreases in an approximately exponential manner $\sim e^{-0.8\zeta t}$ and potentially exceeds the earlier minimum. On the other hand, the width of the density $\sigma_\text{min}$ decreases and eventually converges to $\sigma_{\text{LLL}} = l_B/\sqrt{2}$, indicating the occupation of the lowest Landau level. This is confirmed by the average Landau level occupation $n_a = \langle \tilde{a}^\dagger \tilde{a} \rangle$, shown by the dot-dashed line in Fig.~\ref{Fig1}(c), which approaches zero as $\sigma_{\text{min}}$ reaches $\sigma_{\text{LLL}}$. Fig.~\ref{Fig1}(d) shows the density distribution $\rho(x,y)$ at $\zeta t=2.5$, with dotted lines marking $\pm\sigma_{\text{LLL}}$.

\begin{figure}[t]
	 \includegraphics[width=0.48\textwidth]{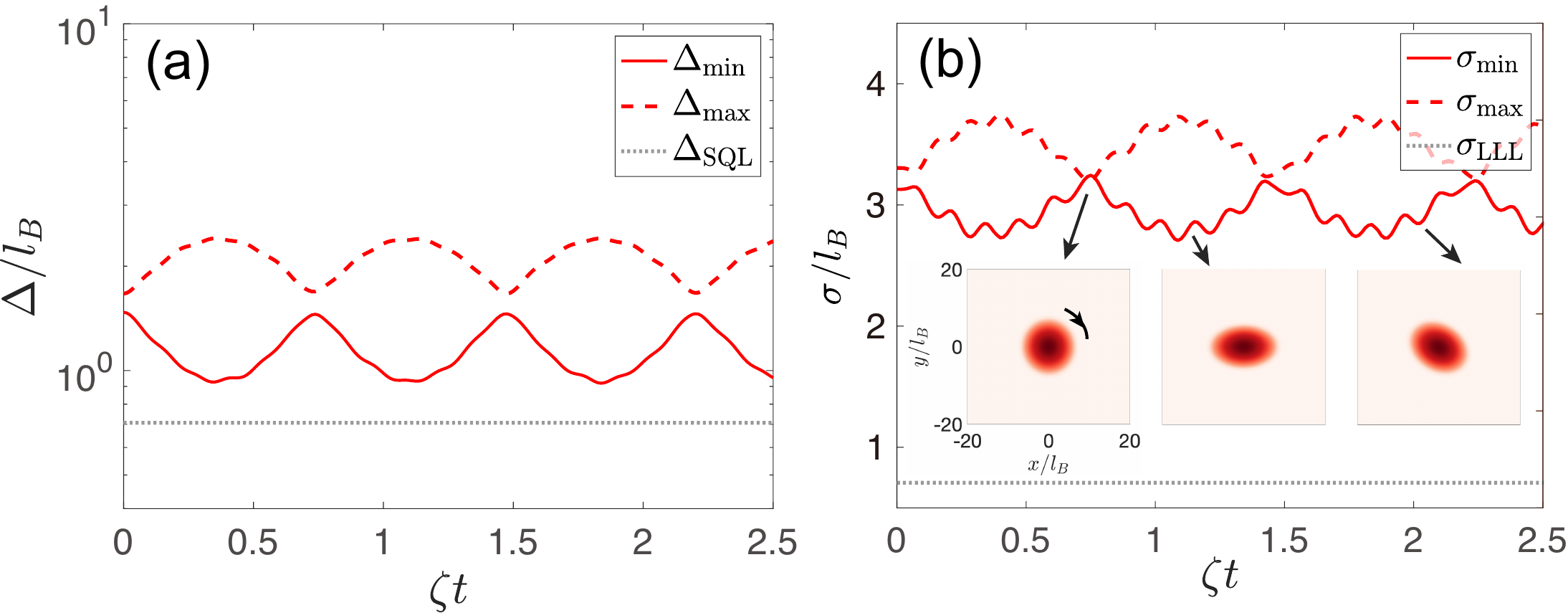}
	\caption{Squeezing dynamics by quenching $\Omega$ to $\omega$ at $t=0$. (a) Variation of $\Delta_\text{min}$ (solid line) and $\Delta_\text{max}$ (dashed line). (b) Evolution of $\sigma_\text{min}$ (solid line) and $\sigma_\text{max}$ (dashed line) in real space. The dotted line represents $\sigma_\text{LLL} = l_B$. Insets show typical density distributions $\rho(x,y)$ at selected temporal moments. All the parameters, as well as the initial state, are the same as those in Fig.~\ref{Fig1}.
	}
	\label{Fig2}
\end{figure}

Now, let us examine what happens when we suddenly quench $\Omega$ from 0 to $\omega$ at $t=0$, with all the other parameters remaining unchanged. The results are discussed in Fig.~\ref{Fig2}. The quench scheme is precisely analogous to that for the single-particle case shown in Sec.~\ref{Model}, but the results are quite different. 
It turns out that the BEC fails to attain geometric squeezing, as evidenced by a near-equilibrium oscillation with frequency $\approx 0.56\omega$ in both the phase space fluctuation $\Delta$ [Fig.~\ref{Fig2}(a)] and the density width $\sigma$ [Fig.~\ref{Fig2}(b)]. As time evolves, the density $\rho(\mathbf{r})$ exhibits clockwise rotation, as illustrated in the insets of Fig.~\ref{Fig2}(b).

The striking disparities in the dynamics between the interacting and non-interacting BECs motivate us to investigate the origin of these differences and propose possible quench-based geometric squeezing protocols for interacting BECs that can outperform the ramping scheme shown in Fig.~\ref{Fig1}. These questions will be addressed in the following two sections.

\section{Hydrodynamics} \label{Hydrodynamics}
Here we will use the hydrodynamic approach to gain some insights into the squeezing dynamics of an interacting BEC.

\subsection{Stability Analysis}
We consider the hydrodynamic description of the BEC by rewriting the wave function as
$\psi(\mathbf{r},t) = \sqrt{\rho(\mathbf{r},t)} e^{i[\theta(\mathbf{r},t) - \mu t]}$,
where $\rho(\mathbf{r},t)$ and $\theta(\mathbf{r},t)$ denote the density field and the phase field, respectively, and $\mu$ is the chemical potential.
The GP equation [Eq.~(\ref{GPE})] can then be rewritten as
\begin{eqnarray}
\dot{\rho}&=&-\nabla \cdot(\rho \mathbf{v}), \label{neq} \\
m \dot{\mathbf{v}}&=&-\nabla\left[\delta \mu+\frac{1}{2} m v^2\right], \label{veq}
\end{eqnarray}
where the first and second lines are coupled equations for the density field $\rho$ and velocity field $\mathbf{v}$. Here, 
\begin{eqnarray}
\mathbf{v}&=& (v_x,v_y) = \frac{1}{m} \left(\nabla \theta -\mathbf{A} \right) \label{v}, \\
\delta \mu &=&g \rho-\frac{1}{2 m \sqrt{\rho}}\nabla^2 \sqrt{\rho} +\tilde V_0^\text{eff}(\mathbf{r})-\mu, \label{delta_n}
\end{eqnarray}
with $\mathbf{A} = (-m \Omega y, m \Omega x,0)$ being the gauge potential, and 
\begin{equation}
\tilde V_0^\text{eff}(\mathbf{r}) = \frac{m (\omega^2 - \Omega^2)}{2} (x^2+y^2) + \frac{\varepsilon m \omega^2}{2} (x^2-y^2).
\end{equation}
Eq.~(\ref{v}) indicates that, in the rotating frame, the velocity field is associated with the mechanical momentum, depending both on the phase gradient and the vector potential $\mathbf{A}$.

The steady-state solutions of the hydrodynamic equations satisfy $\dot{\rho} = \dot{\mathbf{v}} = 0$. For a non-interacting BEC with $g=0$, the non-vortex steady states are Gaussian with $\rho(\mathbf{r})=N\left(m \sqrt{\tilde{\omega}_x \tilde{\omega}_y}/ \pi\right)^{1 / 2} \exp \left[-m\left(\tilde{\omega}_x x^2+\tilde{\omega}_y y^2\right)\right]$ and 
\begin{equation}
	\mathbf{v}_0=\alpha \nabla x y-\frac{\mathbf{A}}{m},
	\label{v0}
\end{equation}
with ${\omega}_{x,\text{eff}}$ and ${\omega}_{y,\text{eff}}$ being the effective trapping frequencies. Inserting Eq.~(\ref{v0}) into Eq.~(\ref{veq}) leads to 
\begin{equation}
	\begin{aligned}
	& {\omega}_{x,\text{eff}}^2=(1+\varepsilon) \omega^2+\alpha^2-2 \alpha \Omega, \\
	& {\omega}_{y,\text{eff}}^2=(1-\varepsilon) \omega^2+\alpha^2+2 \alpha \Omega,
	\end{aligned}
	\label{omega_eff}
\end{equation}
based on which one can derive [according to Eq.~(\ref{neq})] 
\begin{equation}
\alpha=-\Omega \,\frac{\omega_{x,{\text{eff}}}-\omega_{y,{\text{eff}}}}{\omega_{x,{\text{eff}}}+\omega_{y,{\text{eff}}}}.
\label{alpha_non_int}
\end{equation}
On the other hand, for the interacting BEC in the Thomas-Fermi regime (with $g\gg0$), the non-vortex steady states are given by $\rho_0=g^{-1}\left[\mu-m\left({\omega_{x,\text{eff}}^{2}} x^2+{{\omega}_{y,\text{eff}}^2} y^2\right)/2\right]$ for $r\le R_\text{TF}$, with $\mu = (g N m {\omega}_{x,\text{eff}} {\omega}_{y,\text{eff}}/\pi)^{1/2}$ and $R_\text{TF} = (4 g/\pi m \omega^2)^{1/4}$ being the Thomas-Fermi radius. Now, Eq.~(\ref{omega_eff}) remains valid, but $\alpha$ takes a different form as
\begin{equation}
	\alpha=-\Omega \,\frac{\omega_{x,{\text{eff}}}^2-\omega_{y,{\text{eff}}}^2}{\omega_{x,{\text{eff}}}^2+\omega_{y,{\text{eff}}}^2}.
	\label{alpha_int}
\end{equation}

It is also convenient to introduce a dimensionless parameter
\begin{equation}
\gamma = \frac{{\omega}_{x,\text{eff}}^2}{{\omega}_{y,\text{eff}}^2} = \frac{(1+\varepsilon) \omega^2+\alpha^2-2 \alpha \Omega}{(1-\varepsilon) \omega^2+\alpha^2+2 \alpha \Omega}
\end{equation}
 to characterize the anisotropy of $\rho_0$. Stable solutions require ${\omega}_{x,\text{eff}}$ and ${\omega}_{y,\text{eff}}$ to be real and non-negative, rendering $\gamma \ge 0$. Since the isodensity lines of $\rho_0$ form ellipses, $\gamma > 1$ ($\gamma <1$) means the elliptical long axis aligns along the $y$-axis ($x$-axis), and $\gamma=1$ corresponds to an isotropic density profile. 
 
\begin{figure}[t]
	 \includegraphics[width=0.48\textwidth]{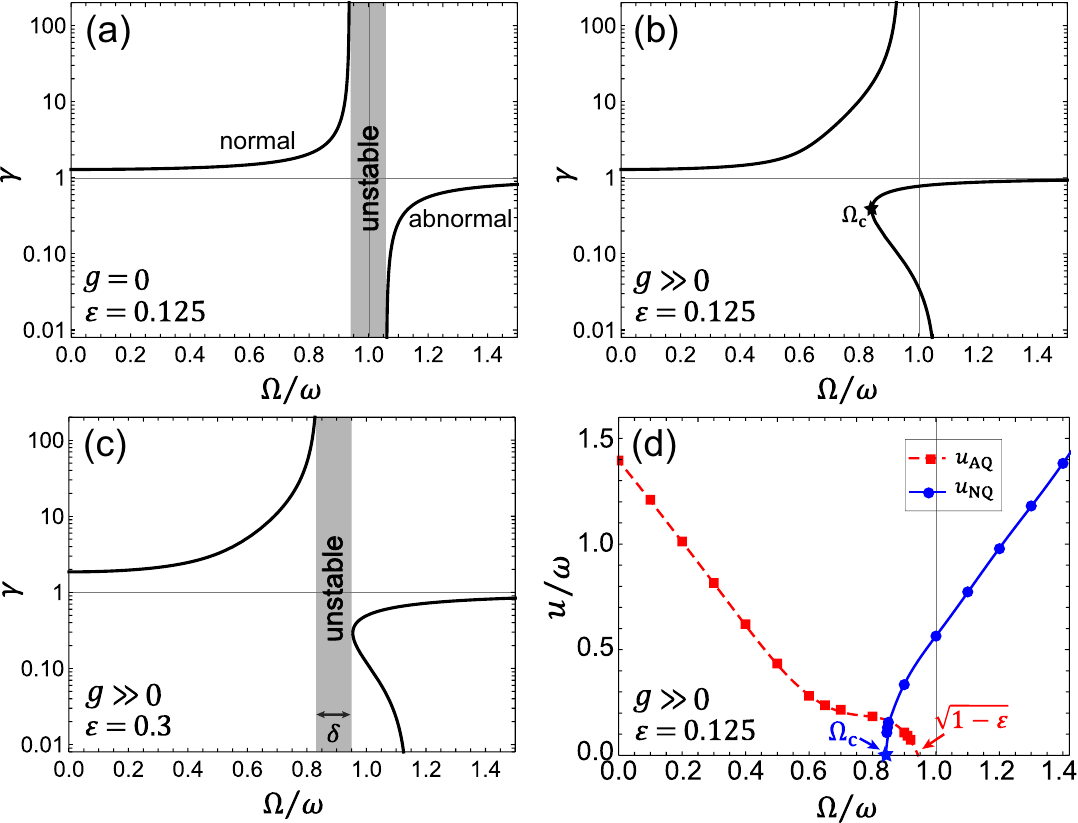}
	\caption{(a)-(c) Steady solutions for the non-interacting $g=0$ and interacting BECs $g\gg 0$. The solid line denotes the dependence of the steady-state anisotropy $\gamma$ on $\Omega$. The shading areas denote the unstable window in which no vortexless steady states can be found. The star in the pannel (b) marks the turning point of the abnormal branch at $\Omega = \Omega_c \approx 0.84 \omega$. (d) The quadrupole mode's frequency $u$ as a function of $\Omega$ for $\varepsilon = 0.125$.
}
	\label{Fig3}
\end{figure}

 For isotropic trapping potential with $\varepsilon=0$, regardless of whether the interaction is turned on or not, there always exists an isotropic steady state with $\omega_{x,\text{eff}} = \omega_{y,\text{eff}}$ such that $\gamma=1$. However, for $\varepsilon \neq 0$, we need to numerically solve Eqs.~(\ref{omega_eff})-(\ref{alpha_int}). In Fig.~\ref{Fig3}(a) and Fig.~\ref{Fig3}(b), we set $\varepsilon = 0.125$ and show the dependence of $\gamma$ on $\Omega$ for the non-interacting and interacting BECs, respectively. There exist two branches of steady-state solutions, the normal and the abnormal (also called the overcritical \cite{Recati2001}) branches, as illustrated in Fig.~\ref{Fig3}(a). The normal branch is located within the region $\Omega \in [0,\sqrt{1-\varepsilon}\omega]$, characterized by $\gamma > 1$. The abnormal branch occupies the region $\Omega\in [\sqrt{1+\varepsilon}\omega, \infty)$, satisfying $\gamma \le 1$, which indicates that the long axis of $\rho_0$ is oriented along the $x$-axis, opposite to the anisotropy direction of the external potential.

Comparing panels (a) and (b) in Fig.~\ref{Fig3}, the major difference between the non-interacting and interacting cases lies in the monotonicity of $\gamma$ along the abnormal branch. Specifically, for the non-interacting BEC with $g=0$, $\gamma$ changes monotonically with $\Omega$. There exists an unstable window (the shading area) within the regime $\Omega \in [\sqrt{1-\varepsilon} \omega,\sqrt{1+\varepsilon}\omega]$, in which no vortexless steady states can be found.
For the interacting BEC with $g\gg0$, the terminating frequency for the normal/abnormal branch remains at $\sqrt{1\mp\varepsilon}\omega$, but $\gamma$ is non-monotonic and exhibits a turning point at $\Omega = \Omega_c$ (marked by the star), which indicates the minimal value of $\Omega$ that can be reached by the abnormal branch. The turning point $\Omega_c$ is determined by the equation
\begin{equation}
\frac{\varepsilon^2 \Omega_c^2}{2 \omega^2}+\frac{2\left(\omega^2-2 \Omega_c^2\right)^3}{27 \omega^6}=0. \label{omegaceq}
\end{equation}
The value of $\Omega_c$ increases with $\varepsilon$, and for small values of $\varepsilon$, it behaves like 
\begin{equation}
\Omega_c \approx \frac{\left(4+3 \varepsilon^{2 / 3}\right) \omega}{4 \sqrt{2}}.
\label{Omegac}
\end{equation}
For $\varepsilon = 0.125$, we have $\Omega_c\approx 0.84\omega < \sqrt{1-\varepsilon}\omega$. Within the regime $\Omega \in [\Omega_c, \sqrt{1+\varepsilon}\omega]$, two steady abnormal states coexist, with the upper state exhibiting much smaller anisotropy than the lower one. Therefore, for such a small value of $\varepsilon$, the interacting BEC does not exhibit an unstable window, in contrast to the non-interacting case. 

Now, it is time to provide a physical picture underlying the squeezing dynamics presented in the previous section. The quench approach of squeezing uses the dynamical instability by quenching the isotropic ground state of BEC at $\Omega=0$ into the unstable window. For the non-interacting BEC, 
the critical frequency $\Omega=\omega$ is within the unstable window. However, the atomic interaction $g$ modifies the BEC's stability condition such that there is no unstable window. At $\Omega=\omega$, the abnormal branch is stable. Therefore, the dynamical oscillation observed in Fig.~\ref{Fig2} should correspond to a certain type of collective excitation of the abnormal state.

However, the ramping approach adopted by the experiment \cite{Fletcher2021} (shown in Fig.~\ref{Fig1}) is quasi-adiabatic. It generates squeezing by first ramping up $\Omega$, allowing the system to follow the normal branch shown in Fig.~\ref{Fig3}(b). Then, as $\Omega \approx \sqrt{1-\varepsilon}\omega$, the collective quadrupole mode's frequency $u_\text{AQ}$ approaches zero [see Fig.~\ref{Fig3}(d) and the discussion below], which breaks the adiabatic theorem. This indicates that the ramping method and the quench method produce squeezing through completely different mechanisms. We would explore viable quench schemes for interacting BECs in Sec.~\ref{RPS}.

\subsection{Quadrupole Mode}
To get a detailed understanding of the oscillating frequency ($\approx 0.56 \omega$) observed in Fig.~\ref{Fig2}, we now investigate the collective excitations. Linearizing the density field $\rho$ and the velocity field $\mathbf{v}$ of Eqs.~(\ref{neq}) and (\ref{veq}) around the steady-state solution $\rho_0$ and $\mathbf{v}_0$, we obtain
\begin{equation}
\begin{aligned}
\delta \dot{\rho} & =-\nabla \cdot[\rho_0 \delta \mathbf{v}+\delta \rho \mathbf{v}_0], \\
\dot{\delta} \mathbf{v} & =-\nabla\left(\frac{g}{m} \delta \rho+\mathbf{v}_0 \cdot \delta \mathbf{v}\right),
\end{aligned}
\end{equation}
where second- and higher-order quantum fluctuations have been neglected. Assuming the fluctuations are periodic, such that $\delta \rho(\mathbf{r},t) = e^{-i u t} \delta \rho(\mathbf{r})$ and $\delta \mathbf{v}(\mathbf{r},t) = e^{-i u t} \delta \mathbf{v}(\mathbf{r})$ with $u$ being the oscillating frequency of collective modes, we obtain the coupled linear equations
\begin{equation}
\begin{aligned}
u \delta \rho & =-i \nabla \cdot[\rho_0 \delta \mathbf{v}+\delta \rho \mathbf{v}_0], \\
u {\delta} \mathbf{v} & =- i \nabla\left(\frac{g}{m} \delta \rho+\mathbf{v}_0 \cdot \delta \mathbf{v}\right).
\end{aligned}
\label{deltaEqs}
\end{equation}
Note that the conjugate fields, $\delta \rho^*$ and ${\delta} \mathbf{v}^*$, also satisfy Eq.~(\ref{deltaEqs}) but with a frequency of $-u$. Hence, physical collective modes with real density fluctuations should be
\begin{equation}
e^{-i u t} \delta \rho + e^{i u t} \delta \rho^* \propto \cos(u t)\text{Re}(\delta \rho) + \sin(u t)\text{Im}(\delta \rho),
\label{ReIm}
\end{equation}
where $\text{Re}(\cdot)$ and $\text{Im}(\cdot)$ denote the real and imaginary parts, respectively. Each solution $(\delta \rho,\delta \mathbf{v})$ with a real $u$ represents a dynamically stable collective mode. Otherwise, the mode is dynamically unstable.

\begin{figure}[t]
	 \includegraphics[width=0.48\textwidth]{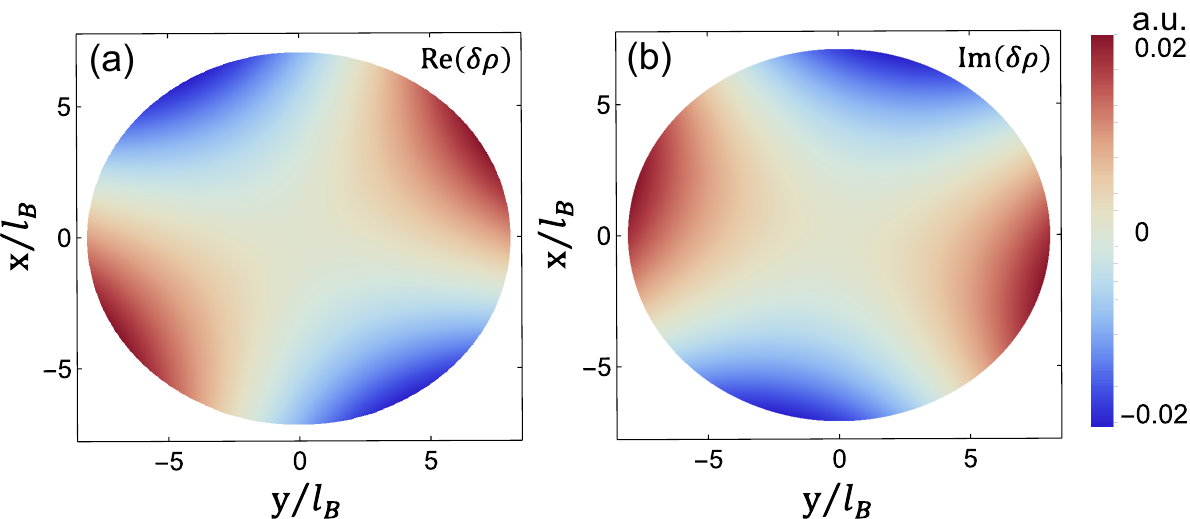}
	\caption{(a) and (b) respectively illustrate the real and imaginary parts of the density field of the AQ mode, with $\Omega = \omega$ and $\varepsilon = 0.125$.
}
	\label{Fig4}
\end{figure}

We numerically solve Eq.~(\ref{deltaEqs}), and find that the quadrupole mode is responsible for the aforementioned oscillation shown in the quench protocol for interacting BEC (see Fig.~\ref{Fig2}). Other collective modes, such as the monopole mode, are less relevant.
In Fig.~\ref{Fig3}(d), we plot the frequency of the quadrupole mode for both the normal branch $u_\text{NQ}$ (dashed line with squares) and the upper abnormal branch $u_\text{AQ}$ (solid line with dots), with $\varepsilon = 0.125$ being fixed. The quadrupole mode of the lower abnormal branch turns out to be dynamically unstable. At the critical rotation speed $\Omega = \omega$, the quadrupole frequency of the upper abnormal state is $u_\text{AQ} \approx 0.564\omega$, which is exactly the oscillating frequency observed in Fig.~\ref{Fig2}. Furthermore, the real and imaginary parts of the density fluctuation $\delta n$ of the AQ mode are shown in Figs.~\ref{Fig4}(a) and (b), respectively. As per Eq.~(\ref{ReIm}), the real and imaginary parts of $\delta \rho$ alternate during evolution, which explains the clockwise density rotation observed in Fig.~\ref{Fig2}(b).

\section{Quench Approach for Interacting BECs} \label{RPS}
From the above discussions, we see that the different stability conditions (in particular, the presence and the lack of unstable window for non-interacting and interacting BECs, respectively) require different approaches to generate squeezing in non-interacting and interacting BECs confined in harmonic traps with small anisotropy. For the former, a simple quench approach suffices, whereas a ramping approach is required for the latter. 

Can we then create an unstable window for interacting BECs? If the answer is yes, then one should be able to generate squeezing in interacting BECs using a quench protocol. In order for such an unstable window to exist, we need $\Omega_c$ (the minimum rotation frequency for the abnormal branch) to exceed $\sqrt{1-\varepsilon} \omega$ (the maximum rotation frequency for the normal branch). Eq.~(\ref{Omegac}) shows that $\Omega_c$ increases as the trap anisotropy $\varepsilon$. Hence, this condition might be satisfied for a large enough $\varepsilon$. It turns out that this expectation is correct. Numerically solving Eq.~(\ref{omegaceq}) shows that $\Omega_c >\sqrt{1-\varepsilon} \omega $
for $\gtrsim 0.2 \omega$. 
An example is shown in Fig.~\ref{Fig3}(c), in which we show the steady-state results for the case of $\varepsilon = 0.3$, where the unstable window marked by the shaded area approximately spans $\Omega \in [0.84\omega, 0.95\omega]$.

\begin{figure}[t]
    \includegraphics[width=0.49\textwidth]{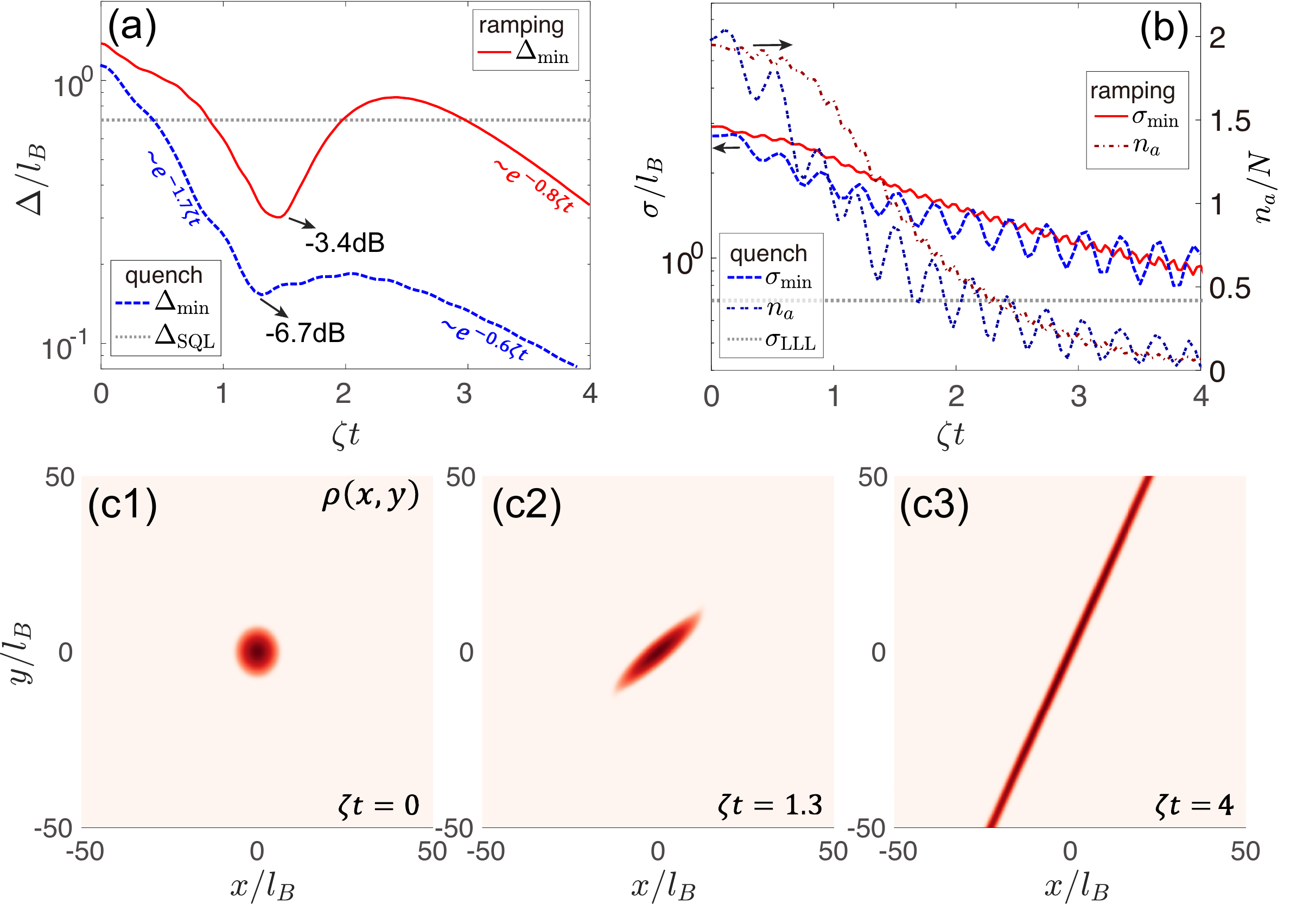}
    \caption{Generation of geometric squeezing using the quench scheme for interacting BECs. (a) The blue dashed line depicts the minimal quantum fluctuations $\Delta_\text{min}$ in phase space; the red solid line shows the results of the ramping scheme which has already been shown in Fig.~\ref{Fig1}(b). (b) The blue dashed line represents the evolution of the minimal wave-packet width $\sigma_\text{min}$ (left axis); the double-dotted line indicates the Landau-level occupation $n_a$ (right axis). We also illustrate the results of the ramping scheme using red curves. Arrows indicate the corresponding axes for quantities.
    (c) The real-space density distribution at moments $\zeta t=0$, $1.3$, and $4$, with $\zeta \approx 0.167 \omega$.
    }
    \label{Fig5}
\end{figure}

With the emergence of the unstable window, we can now investigate the quench protocol for the interacting case, with the results presented in Fig.~\ref{Fig5}. In the calculation, we take the initial state to be the non-rotating ground state in a harmonic trap with anisotropy $\varepsilon = 0.3$, and then quench it under the government Hamiltonian with $\Omega = 0.9\omega$ (accordingly $\zeta \approx 0.167 \omega$) which locates inside the unstable window. Panel (a) of Fig.~\ref{Fig5} shows the evolution of the minimal phase-space quantum fluctuations $\Delta_\text{min}$, while panel (b) displays the real-space width $\sigma_\text{min}$ and the Landau-level occupation $n_a$. Typical density distributions $\rho(t)$ at selected times are displayed in panel (c). For comparisons, the results of the ramping approach (previously shown in Fig.~\ref{Fig1} with $\varepsilon=0.125$) are also presented by red curves in Figs.~\ref{Fig5}(a) and (b).

It can be observed in the figures that the quench scheme exhibits clear advantages over the ramping approach. 
In phase space, $\Delta_\text{min}$ also exhibits a two-stage decay, with the first decay being approximately exponential $\sim \exp(-1.7 \zeta t)$, and reaches its minimum at $\zeta t \approx 1.3$, achieving approximately $-6.7$ dB of geometric squeeze, which is significantly stronger than the $-3.4$dB achieved by the ramping method. For $\zeta t > 2.4$, $\Delta_\text{min}$ enters a second stage of decline, scaling with $\sim \exp(-0.6 \zeta t)$. Unlike the ramping approach, $\Delta_\text{min}$ for the quench scheme stays consistently below the SQL from the onset of squeezing.

In real space, both $\sigma_\text{min}$ and $n_a$ decrease more rapidly in the short term compared to the ramping approach. Over longer periods, the BEC fully enters the lowest Landau levels at a rate similar to that of the ramping approach, as shown in Figs.~\ref{Fig5}(b) and (c). 

\begin{figure}[t]
    \includegraphics[width=0.40\textwidth]{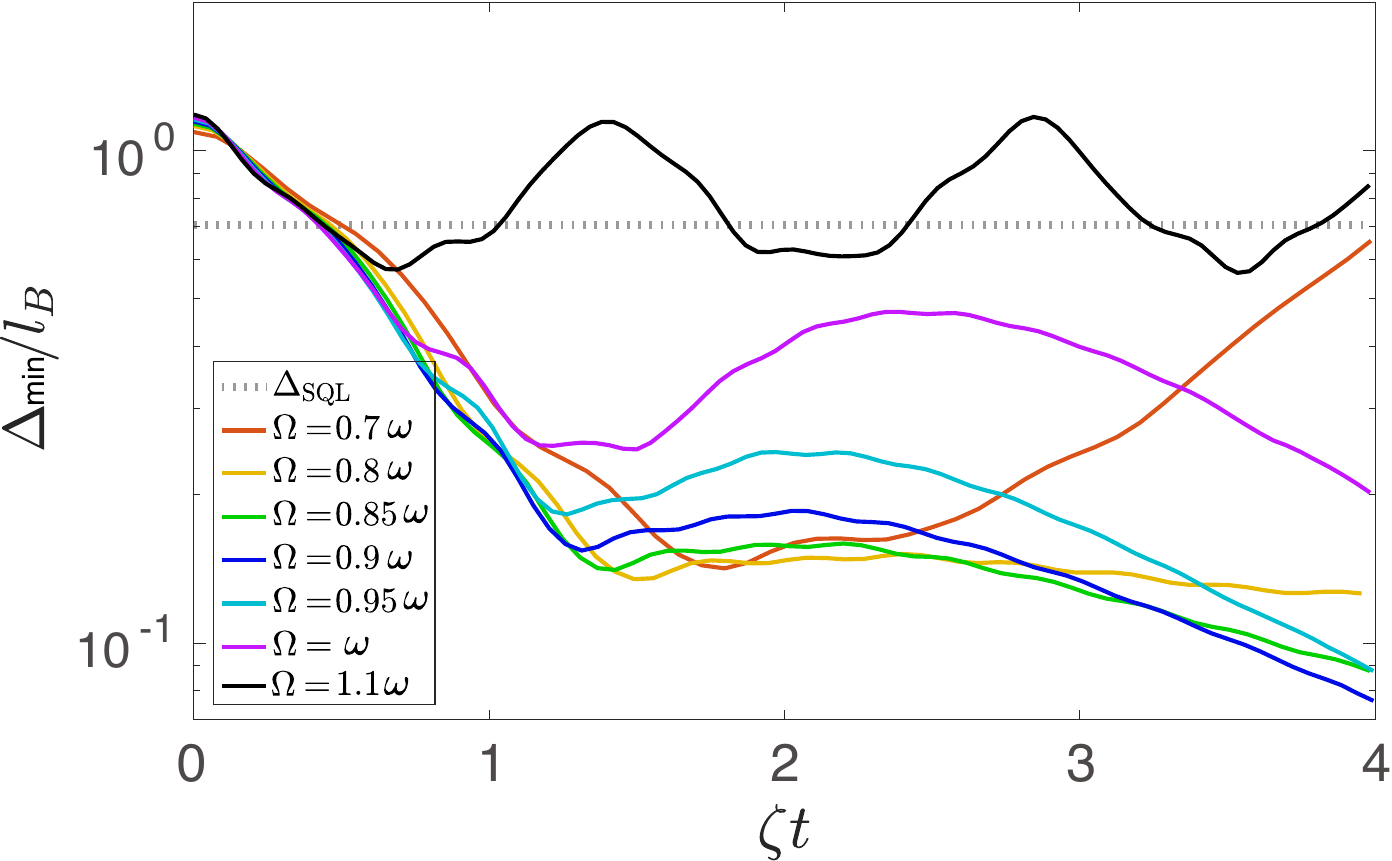}
    \caption{The squeezing dynamics for interacting BECs with fixed $\varepsilon = 0.3$, by quenching from the non-rotating ground state at $\Omega = 0$ to various target $\Omega$. The dotted line indicates the SQL.}
    \label{Fig6}
\end{figure}

In Fig.~\ref{Fig6}, we compare the squeezing dynamics by quenching the non-rotating ground state to different target rotation frequencies $\Omega$, where the case with $\Omega = 0.9 \omega$ has already been shown in Fig.~\ref{Fig5}. The results clearly reveal a two-stage squeezing dynamics for rotation frequencies within the unstable window ($\Omega \in [0.84\omega, 0.95\omega]$). For target frequencies sufficiently outside this unstable window, such as $\Omega = 0.7\omega$ and $\Omega = 1.1\omega$, the BEC cannot achieve a long-time squeeze. This further confirms our stability analysis that effective geometric squeezing requires breaking the BEC's stability. In particular for the case of $\Omega = 1.1\omega$, periodic oscillations reappear (similar to Fig.~\ref{Fig2}(a)), which are again due to collective oscillations of the abnormal branch shown in Fig.~\ref{Fig3}(c).
Within the unstable window, rotation frequencies closer to the center of the window ($\Omega = 0.9\omega$) produce stronger long-term squeezing compared to those near the edges ($\Omega = 0.85\omega$ or $\Omega = 0.95\omega$).

\section{Discussion} \label{Discussion}
\subsection{Moment of Inertia}
Generally in rotating systems, the moment of inertia is a crucial quantity characterizing the system's resistance to changes in its rotation. In the current system, the moment of inertia is defined as \cite{Zambelli2001}
\begin{equation}
I = \frac{\langle L_z \rangle}{\Omega},
\end{equation}
where $\langle L_z \rangle$ is the expectation value of the angular momentum. Since the anisotropy $\varepsilon$ in potential $V_0$ breaks the rotational symmetry of the Hamiltonian in the $x$-$y$ plane, the angular momentum $L_z$ is no longer conserved. For the squeezing dynamics discussed above, regardless of whether the system is interacting or not, the angular momentum always increases over time. 

\begin{figure}[t]
	 \includegraphics[width=0.49\textwidth]{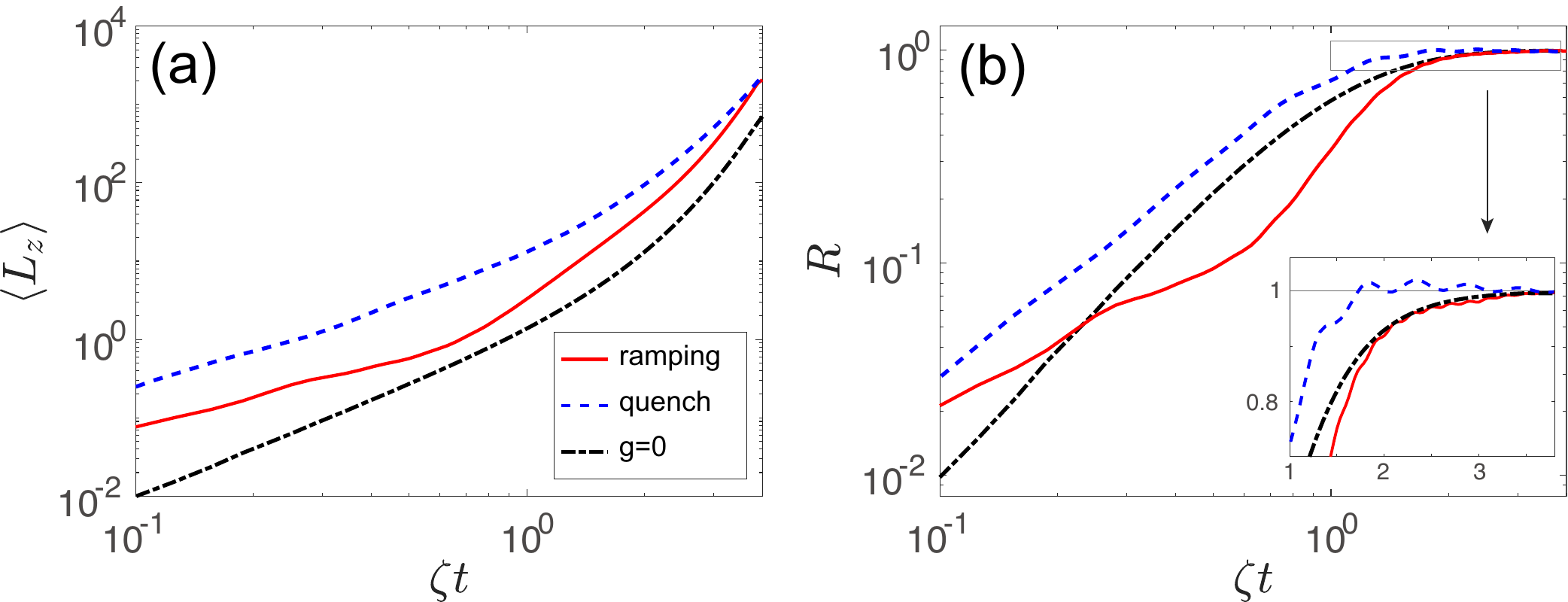}
	\caption{(a) Log-log plot of the angular momentum $\langle L_z \rangle$ versus $t$, where the solid line and dashed line respectively represent the ramping-$\Omega$ approach and the quench dynamics for interacting BECs; the dot-dashed line represents the squeezing dynamics of the non-interacting BEC. (b) Log-log plot of the ratio $R$ between the BEC's rotational inertia and the classical rigid-body value. The inset magnifies the portion within the black box and uses normal axes.
}
	\label{Fig7}
\end{figure}

Particularly for the non-interacting BEC with $g=0$, the wave function of the squeezed state in the coordinate representation is given by \cite{Fletcher2021,Chen2025}
\begin{equation}
\begin{aligned}
\psi(\mathbf{r},t) &= \langle x,y|0,S\rangle \\
&= \frac{\exp \left(-\frac{x^2+y^2 + i(x+i y)^2 \tanh (\zeta t)}{4 l_B^2}\right)}{\sqrt{2 \pi \cosh (\zeta t)} l_B},
\end{aligned}
\end{equation}
based on which one can obtain
\begin{equation}
\langle L_z \rangle = \sinh^2(\zeta t),
\end{equation}
which leads to 
\begin{equation}
I = \frac{\sinh^2(\zeta t)}{\omega}.
\end{equation}
For small $\zeta t$, the leading order of $\langle L_z \rangle$ is $(\zeta t)^2$, so $\langle L_z \rangle$ exhibits a power-law growth as shown by the dot-dashed line in Fig.~\ref{Fig7}(a). For a long time, higher-order terms become relevant, resulting in nonlinear growth in the log-log plot. In Fig.~\ref{Fig7}(a), we also present the numerical results of $\langle L_z \rangle$ for the ramping approach (denoted by the solid line, corresponding to the case in Fig.~\ref{Fig1}) and the quench dynamics (denoted by the dashed line, corresponding to the case in Fig.~\ref{Fig5}). For a short time, $\langle L_z (t)\rangle$ of the quench scheme exhibits a similar tendency with that of the non-interacting BEC, and is significantly faster than the ramping approach; whereas over a long time, the behavior of $\langle L_z (t)\rangle$ for the quench and the ramping approaches align more closely with each other.

Another relevant quantity of interest is the ratio $R$ of the moment of inertia to the value of rigid body, i.e.,
\begin{equation}
R = \frac{I}{I_\text{rig}},
\end{equation}
where
\begin{equation}
I_\text{rig} = m \langle r^2 \rangle = m \int d^2 r r^2 \rho(\mathbf{r})
\end{equation}
is the inertia moment of a rigid body. The conventional interacting BECs in equilibrium feature $R\ll 1$, which is a direct manifestation of the lack of viscosity and friction for a superfluid within a container. However, the geometric squeeze process represents a crossover of the superfluid toward the rigid body, thus $R$ would increase from zero to one. 

For $g=0$, $I_\text{rig}$ and $R$ have the exact form of
\begin{equation}
I_\text{rig} = 2 m l_B^2 \cosh^2(\zeta t),
\end{equation}
and
\begin{equation}
R = \tanh^2(\zeta t).
\end{equation}
Due to the properties of $\tanh$ function, $R$ exhibits a monotonic increase and saturates at one, as shown by the dot-dashed line in Fig.~\ref{Fig7}(b). The figure also presents $R(t)$ for ramping-$\Omega$ evolution and the quench scheme for the interacting BEC with $g\gg 0$. An interesting observation is that, in the quench scheme, the BEC's moment of inertia can slightly exceed that of a rigid body and exhibits a non-monotonicity. This is evidenced in the insets of Fig.~\ref{Fig7}(b) that $\langle R (t)\rangle > 1$ for $\zeta t \gtrsim 1.8$, peaking around $\zeta t \approx 2$, and then slowly falling back to one.

\subsection{Experiment Considerations}
The realization of our quench schemes does not require more complex experimental techniques than those already used in the experiment \cite{Fletcher2021}: The BEC is initially prepared in the non-rotating ground state with anisotrpic trap $\varepsilon$, followed by a rapid ramp up of $\Omega$ to the target values, and then the system is allowed to evolve until measurement. In the real $x$-$y$ space, the density distribution $\rho(\mathbf{r})$ of the BEC can be directly measured through \textit{in-situ} absorption imaging, and its width $\sigma_\text{min}$ can reflect the extent to which the BEC enters the lowest Landau level. 

A potential challenge lies in detecting the squeezing parameters in phase space, as $X$ (or $Y$) is composed of both the coordinate $\mathbf{r}$ and the momentum $\mathbf{p}$, making it impossible to measure the quantum fluctuations independently in the coordinate or momentum space. A possible solution might be to reconstruct the quasi-probability distribution from the density distribution $\rho$. For example, if the Wigner function of the cyclotron $\xi$-$\eta$ phase space is Gaussian, then $\rho$ represents the Husimi-Q function in the guiding-center space \cite{Fletcher2021}. 
However, for interacting BECs whose Wigner function deviates from Gaussian, this approach may have limitations. Therefore, state tomography in geometric phase spaces can serve as a promising direction for future research.

\section{Conclusion} \label{Conclusion}
In this paper, we have systematically investigated the dynamical generation of geometric squeezing in interacting BECs. Our work demonstrates that the quench scheme for squeezing generation in both non-interacting and interacting BECs can be described within a unified theoretical framework, with the key insight being the disruption of the BEC's dynamical stability. For non-interacting BECs, the system is always unstable near the critical rotation frequency. However, atomic interactions significantly alter the system's stability, rendering the quench protocol failure at the critical rotation frequency. By appropriately modifying the superfluid stability criteria via the change of trap anisotropy, we enable efficient generation of geometric squeezing using the quench approach. We also demonstrate the rigidification process of the BEC during dynamical squeezing, showing how the moment of inertia evolves from a superfluid toward a rigid body.

\begin{acknowledgments}
L. C. acknowledges support from the NSF of China (Grant No. 12174236) and from the fund for the Shanxi 1331 Project. H.P. acknowledges support from the NSF (Grant No. PHY-2207283) and the Welch Foundation (Grant No. C-1669).
\end{acknowledgments}

\end{document}